\documentclass[twocolumn,showpacs,preprintnumbers,amsmath,amssymb,superscriptaddress]{revtex4}
\usepackage{graphicx}
\usepackage{amsmath}
\usepackage{times}
\usepackage{color}

\begin{document}
\title{Effects of inhomogeneous influence
of individuals on an order-disorder transition in opinion
dynamics}

\author{Jian-Yue Guan}
\thanks{guanjr03@lzu.cn} \affiliation{Institute of Theoretical
Physics, Lanzhou University, Lanzhou Gansu, 730000, China}

\author{Zhi-Xi Wu}
\thanks{wupiao2004@yahoo.com.cn} \affiliation{Institute of
Theoretical Physics, Lanzhou University, Lanzhou Gansu, 730000,
China} \affiliation{Department of Electronic Engineering, City
University of Hong Kong, Kowloon, Hong Kong, China}
\author{Ying-Hai Wang}
\thanks{yhwang@lzu.edu.cn}\affiliation{Institute of Theoretical
Physics, Lanzhou University, Lanzhou Gansu, 730000, China}
\date{Received: date / Revised version: date}
\begin{abstract}
We study the effects of inhomogeneous influence of individuals on
collective phenomena. We focus analytically on a typical model of
the majority rule, applied to the completely connected agents. Two
types of individuals $A$ and $B$ with different influence activity
are introduced. The individuals $A$ and $B$ are distributed
randomly with concentrations $\nu$ and $1-\nu$ at the beginning
and fixed further on. Our main result is that the location of the
order-disorder transition is affected due to the introduction of
the inhomogeneous influence. This result highlights the importance
of inhomogeneous influence between different types of individuals
during the process of opinion updating.
\end{abstract}
 \maketitle

In recent years, a large class of interdisciplinary problems has
been successfully studied with methods of statistical physics, in
particular those related to the characterization of the collective
social behavior of individuals, such as opinion formation
\cite{Galam1,Sznajd}, the spreading of rumor or disease
\cite{Galam2,Watts,Boguna}, the language dynamics \cite{Dall},
etc. The study of opinion dynamics has become a main stream of
research in physics \cite{Kvapivsky,Mobilia,Helbing,Galam3}.
Processes of opinion formation are usually modelled as simple
collective dynamics in which the agents update their opinions
following local majority \cite{Galam1,Kvapivsky} or imitation
\cite{Frachebourg}. In most of these models, agents are located on
the nodes of a graph and endowed with a finite number of available
states, e.g., two states - spin up and spin down. Several works
have revealed that a given model may exhibit very different (even
qualitatively) behaviors depending on its underlying topologies
\cite{Boguna}. Recently, Lambiotte has studied the effect of
degree dispersity on an order-disorder transition
\cite{Lambiotte}.

The heterogeneity of individuals may be an important factor in
opinion \cite{Galam4} or other dynamics \cite{Szolnoki}. In Ref.
\cite{Szolnoki}, Szolnoki and Szab\'{o} have introduced the
effects of inhomogeneous activity of teaching to prisoner's
dilemma game. In their model, two types of players that have
different teaching activities, which characterizing the
master-follower asymmetry between two neighboring players are
taken explicitly into account during the strategy adoption
mechanism. It was found that the introduction of the inhomogeneous
activity of teaching can remarkably enhance the evolution of
cooperation \cite{Szolnoki}. It is natural to consider that the
influence between different types of individuals may be different.
We think that it will be interesting to introduce different types
of people to the opinion dynamics. One can think of a system
consisting of two types of people (just like old and young, or
attractive and repulsive, individuals in some communities)
\cite{Szolnoki}. In this paper, we consider the opinion dynamics
of a system containing two types of individuals ($A$ and $B$). Our
motivation is to explore how the inhomogeneous influence between
two types of individuals affects the order-disorder transition.
For this purpose, a variant of the majority rule (MR) model
introduced by Lambiotte \cite{Lambiotte} will be considered by
assuming only two possible values of $\omega_{xy}$, which
characterizing the influence probability between two types of
people.

Let us first introduce our opinion dynamics model. The population
is composed of $N$ individuals, each of them endowed with an
opinion $o_{i}$ that can be $\alpha$ or $\beta$. Two types of
individuals ($n_{x}=A$ or $n_{x}=B$) are distributed randomly on
the nodes of the network. The concentration of individuals $A$ and
$B$ are denoted by $\nu$ and $(1-\nu)$, respectively. At each time
step, one of the individuals is randomly selected. Then one of the
following two processes may take place. $(i)$ With probability
$q$, the selected node $s$ randomly changes its opinion,
\begin{eqnarray}
o_s \rightarrow \alpha  & {\rm with ~probability~} 1/2,\cr o_s
\rightarrow \beta  & {\rm with ~probability~}1/2.
\end{eqnarray}
$(ii)$ With probability $1-q$, two neighboring nodes of $s$ are
selected and the three individuals in this triplet update their
opinions depending on what types they belong to. First, we define
the influence probability between every two neighboring
individuals as $\omega_{xy}$. When we update the states of the
selected three individuals, if they belong to the same type and
there is one individual whose opinion is different from the other
two individuals', then the individual whose opinion is in minority
changes his opinion with probability $1.0$. Thus, we can obtain
$\omega_{AA}+\omega_{AA}=1.0$ if the three individuals all belong
to the type $A$, and $\omega_{BB}+\omega_{BB}=1.0$ if they all
belong to the type $B$. So, we define the influence probability
between two individuals belonging to the same type as $0.5$, i.e.,
$\omega_{AA}=\omega_{BB}=0.5$. Similarly, we can define the
influence probability between two individuals belonging to two
different types as $\omega_{AB}=\omega_{BA}=\omega$. For the sake
of simplicity, we assume that the influences between individuals
in the same type are always stronger or equal to that in two
different types, i.e., $0\leq \omega\leq0.5$. Thus, if this three
- individual triplet consists of two types of people, for example,
two $A$ and one $B$, then the individual $x$ whose opinion is in
minority changes his opinion with probability $2\omega$ if
$n_{x}=B$ and with probability ($0.5+\omega$) if $n_{x}=A$. It is
straightforward to consider the case of two $B$ and one $A$.
Evidently, if $\omega=0.5$ this model is equivalent to a
homogeneous system studied in Ref. \cite{Lambiotte}.

The parameter $q$ involved in the model measures the competition
between individual choices and neighboring interactions, i.e., the
larger the $q$ is, the more random the system would be; on the
contrary, for smaller $q$, the opinion of individuals would become
more homogeneous. In the case $q = 0$ and $\omega=0.5$, it is well
known that the system asymptotically reaches global consensus
where all nodes share the same opinion \cite{Kvapivsky}. In the
other limiting case $q = 1$, the opinions of individuals are
purely random and the average (over the realizations of the random
process) number of individuals with opinion $\alpha$ at time $t$,
denoted by $A_{t}$, goes to $N/2$ for large $t$. In the following,
we will investigate how the inhomogeneous degree $\nu$ of the
system and the influence probability $\omega$ between different
types of individuals affect the order-disorder transition.
\begin{figure}[hpbt]
\begin{center}
{\includegraphics[width=8.8cm]{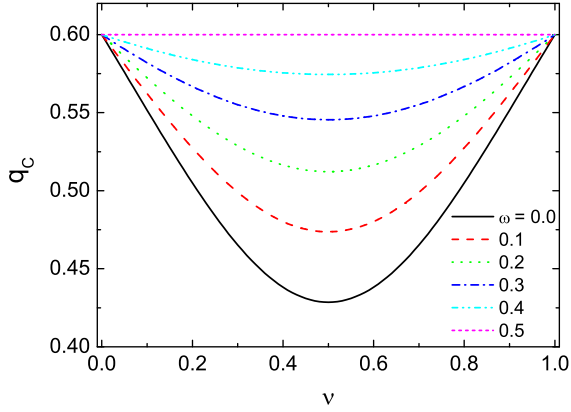}} \caption{(Color
online) The analytical results of the critical values $q_C$ as a
function of $\nu$ for different values of influence probability
$\omega$ (see the plot for detailed values).} \label{fig1}
\end{center}
\end{figure}

In the present model, we assume that the network of individuals is
highly connected and homogeneous, i.e., each individual links to
all other members in the network. In that case, the mean-filed
rate equation for $A_{t}$ reads
\begin{equation}
A_{t+1} = A_{t}+q(\frac{1}{2}-a)+(1-q)W,
\end{equation}
where $a_{t} = A_{t}/N$ is the average proportion of nodes with
opinion $\alpha$, and $W$ is the total contribution to the
evolution of $A_{t}$ due to neighboring interactions. The term
proportional to $q$ accounts for the random flips and the last
term for local majorities. In eq. (2), the total contribution $W$
is
\begin{equation}
W = W_{1} + W_{2} + W_{3} + W_{4},
\end{equation}
where the four terms are the contribution to the evolution of
$A_{t}$ for the cases of three nodes $A$, three nodes $B$, two
nodes $A$ and one node $B$, and one node $A$ and two nodes $B$,
respectively. The probability for two nodes $\alpha$ ($\beta$) and
one node $\beta$ ($\alpha$) to be selected is $3a^{2}(1-a)$
[$3a(1-a)^{2}$], so that

\begin{eqnarray}
W_{1} &=& \nu^{3}[3a^{2}(1-a)-3a(1-a)^{2}] \cr &=&
\nu^{3}[-3a(1-3a+2a^{2})],
\end{eqnarray}
\begin{eqnarray}
W_{2} &=& (1-\nu)^{3}[3a^{2}(1-a)-3a(1-a)^{2}] \cr &=&
(1-\nu)^{3}[-3a(1-3a+2a^{2})],
\end{eqnarray}
\begin{eqnarray}
W_{3} &=& 3\nu^{2}(1-\nu)[a^{2}(1-a)(2\omega) +
2a^{2}(1-a)(\frac{1}{2}+\omega) \cr &-&
a(1-a)^{2}(2\omega)-2a(1-a)^{2}(\frac{1}{2}+\omega)] \cr &=&
3\nu^{2}(1-\nu)(4\omega+1)[-a(1-3a+2a^{2})],
\end{eqnarray}
\begin{eqnarray}
W_{4} &=&
3\nu(1-\nu)^{2}[a^{2}(1-a)(2\omega)+2a^{2}(1-a)(\frac{1}{2}+\omega)
\cr &-& a(1-a)^{2}(2\omega)-2a(1-a)^{2}(\frac{1}{2}+\omega)] \cr
&=& 3\nu(1-\nu)^{2}(4\omega+1)[-a(1-3a+2a^{2})].
\end{eqnarray}
From Eqs.(3), (4), (5), (6), and (7), we obtain
\begin{eqnarray}
W = -3a(1-3a+2a^{2})[2\nu^{2}-2\nu+1+4\omega(\nu-\nu^{2})].
\end{eqnarray}
So the evolution equation for $A_{t}$ can be written as
\begin{eqnarray}
A_{t+1} &=& A_{t}+q(\frac{1}{2}-a)+(1-q) \{-3a(1-3a \cr &+&
2a^{2})[2\nu^{2}-2\nu+1+4\omega(\nu-\nu^{2})]\}.
\end{eqnarray}

\begin{figure}[hpbt]
\begin{center}
{\includegraphics[width=8.8cm]{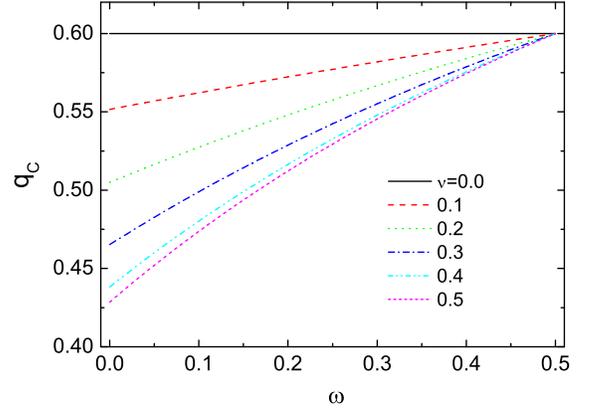}} \caption{(Color
online) The dependence of the critical values $q_C$ on the
influence probability $\omega$ between different types of
individuals for several different values of $\nu$ (see the plot
for detailed values).} \label{fig2}
\end{center}
\end{figure}

It is straightforward to show that $a = 1/2$ is always a
stationary solution of Eq. (9) due to the existence of symmetry.
This is evident after rewriting Eq. (9) for the quantities $\Delta
= A-N/2$ and $\delta = a-1/2$,
\begin{widetext}
\begin{eqnarray}
\Delta_{t+1} = \Delta_{t} +
\frac{\delta}{2}\{3+3(2-4\omega)(\nu^{2}-\nu) -
q[5+3(2-4\omega)(\nu^{2}-\nu)]  -
12(1-q)[(2-4\omega)(\nu^{2}-\nu)+1]\delta^{2}\},
\end{eqnarray}
from which one finds that the symmetric solution $a = 1/2$ ceases
to be stable when $q <
\frac{3+3(2-4\omega)(\nu^{2}-\nu)}{5+3(2-4\omega)(\nu^{2}-\nu)}$,
and that the system reaches the following asymmetric solutions in
that case:
\begin{eqnarray}
a_{-} = \frac{1}{2}- \sqrt{\frac{3+3(2-4\omega)(\nu^{2}-\nu)-
q[5+3(2-4\omega)(\nu^{2}-\nu)]}{12(1-q)[(2-4\omega)(\nu^{2}-\nu)+1]}},
\cr a_{+} = \frac{1}{2}+ \sqrt{\frac{3+3(2-4\omega)(\nu^{2}-\nu)-
q[5+3(2-4\omega)(\nu^{2}-\nu)]}{12(1-q)[(2-4\omega)(\nu^{2}-\nu)+1]}}.
\end{eqnarray}
\end{widetext}
The system therefore undergoes an order-disorder transition at
\begin{eqnarray}
q_{C}(\omega,\nu) =
\frac{3+3(2-4\omega)(\nu^{2}-\nu)}{5+3(2-4\omega)(\nu^{2}-\nu)}.
\end{eqnarray}
Below this value, a collective opinion has emerged because of the
imitation between neighboring nodes. Let us stress that when
$\omega=0.5$ or $\nu=0$ ($\nu=1$), Eqs.(11), respectively,
converge to $a_{-}=0$ and $a_{+}=1$ in the limit $q\rightarrow0$.
We recover the result $q_{C}=3/5$ obtained in Ref.\cite{Lambiotte}
in the limiting case $\omega=0.5$. For the homogeneous
system($\nu=0$ or $\nu=1$), one can also recover the known result
$q_{C}=3/5$ \cite{Lambiotte}. We would resort to pictures to
elucidate that the critical value $q_C$ varies with $\nu$ and
$\omega$ as shown in Eq. (12).

In Fig.\ \ref{fig1}, we show that the critical values $q_C$ of the
order-disorder transition change with the fraction $\nu$ of
individuals $A$ for several different values of $\omega$. We can
find from this figure that, when $\omega<0.5$, the value of $q_C$
decreases monotonously until reaching the minimum value at
$\nu=0.5$, which indicates that the more inhomogeneous the system
is, the smaller the critical value $q_C$ would be. Our first
finding is therefore that the location of the order-disorder
transition depends in a nontrivial way on the inhomogeneous degree
$\nu$ of the system.

\begin{figure}[h]
\begin{center}
{\includegraphics[width=9cm]{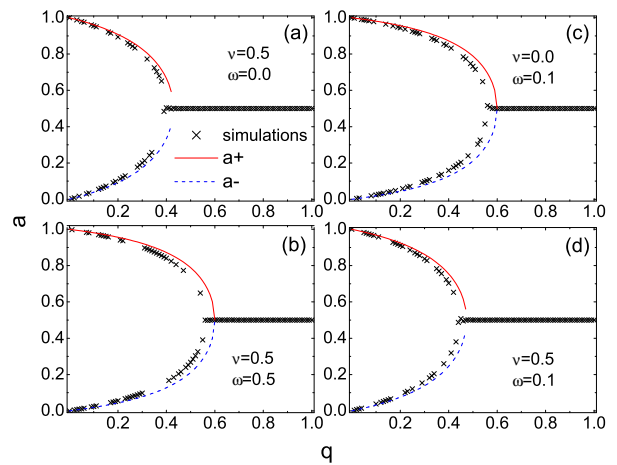}} \caption{(Color online)
The simulation (crosses) and analytical (colored lines) results of
the fraction $a$ of opinion $\alpha$ as a function of $q$ for
different values of $\omega$ and $\nu$: (a) $\nu=0.5$,
$\omega=0.0$; (b) $\nu=0.5$, $\omega=0.5$; (c) $\nu=0.0$,
$\omega=0.1$; (d) $\nu=0.5$, $\omega=0.1$.} \label{fig3}
\end{center}
\end{figure}

Figure \ref{fig2} shows the analytical results for the dependence
of the critical value $q_C$ on the influence probability $\omega$
with fixed value of $\nu$. Due to the symmetry of the system for
two types of individuals, we only consider the cases for
$\nu\leq0.5$. In this figure, $q_C$ increases monotonously with
the parameter $\omega$ until reaching the maximum value $0.6$ at
$\omega=0.5$ for each certain value of $\nu>0$. It indicates that,
the larger the influence probability between two types of
individuals is, the larger the $q_C$ would be. Our second finding
is that the location of the order-disorder transition also depends
on the amount of $\omega$: The smaller the $\omega$ is, the larger
the deviation from the result $0.6$ in the homogeneous system
would be. The effects of the inhomogeneous influence on collective
behaviors may be of some sense in investigating the opinion
dynamics in real social systems.

In order to elucidate the behavior of $a$ below $q_C$, we perform
Eqs. (11) from an analytical point of view in Fig.\ \ref{fig3}. By
construction, the random steps of MR are easy to implement in a
computer simulation. Simulations were carried out for a population
of $N=10000$ individuals located on the sites of the completely
connected network. Other parameters are denoted in Fig.\
\ref{fig3}. We study the key quantity of the fraction $a$ of
opinion $\alpha$ in the steady state. Initially, the two opinions
of $\alpha$ and $\beta$ are randomly distributed among the
individuals with equal probability $1/2$. In simulations, we
denote $\alpha$ and $\beta$ by $+1$ and $-1$, respectively.
Eventually, the system reaches a dynamic equilibrium state. The
simulation results were obtained by averaging over the last $10^4$
Monte Carlo time steps of the total $10^5$. These simulation
results are in very good agreement with Eqs. (11) under the
critical value $q_C$, but it appears as a small difference that
the value of $a$ is smaller when obtained from simulations than
that from analysis near this critical value. This is due to the
finite-size effect.

Finally, we notice that, a transition to a disorder opinion phase
was also obtained in Ref. \cite{Galam5}. The differences between
our work and \cite{Galam5} are that, in Ref. \cite{Galam5}, Galam
considered the contrarians effects on opinion forming, whereas in
the present work we focus on the effects of inhomogeneous
influence of individuals. The similarities are that in both
models, the local majority rule has been used during the process
of opinion updating; and the phenomena the order-disorder
transitions in opinion dynamics all appeared in both Ref.
\cite{Galam5} and our present work. We want to stress that in our
model, only the inhomogeneous influence among individuals (which
is a typical character of many real social systems) is considered,
and one observes rich dynamical phenomena (no other additional
constrain conditions are needed), both the phenomena of the
order-disorder transitions and the changes of the location of
them. The results obtained in \cite{Galam5} may be set in parallel
with recent \lq\lq hung elections \rq\rq as occurred in the $2000$
American presidential elections and that of the $2002$ German
parliamentary elections. Due to the somewhat similarity of the
results obtained by both models, our present work provides an
alternative way to understand the phenomenon of \lq\lq hung
elections \rq\rq \cite{Galam5}.

In summary, we have studied the effects of inhomogeneous influence
of individuals on an order-disorder transition in opinion
dynamics. We mainly considered the majority rule by introducing
two types of people with different influence activity. It was
shown that the location of the order-disorder transition depends
on the inhomogeneous degree $\nu$ of the system and on the
influence probability $\omega$ between different types of
individuals. In social group, the emergence of order means that
there exists a clear cut majority-minority splitting, and in this
phase one can observe polarization of opinions; while in the
disordered phase, there is no opinion dominating with both state
densities equal and no global symmetry breaking. From a social
point of view, our results suggest that it is more difficult to
realize the ordered state in the real world (the value of $q_C$ is
smaller in the inhomogeneous case than that in the homogeneous
case), because most real social systems behave like the
inhomogeneous case. Thus, the inhomogeneous influence of
individuals is a correlated factor in opinion dynamics, which
plays an important role in the opinion spreading and formation in
real systems.

\medskip This work was supported by the National Natural Science
Foundation of China under Grant No. 10775060.

\bibliographystyle{h-physrev3}

\end{document}